# Oriented Bounding Boxes Using Multiresolution Contours for Fast Interference Detection of Arbitrary Geometry Objects

Luis Rivera[1], Vania Estrela[1], Paulo C. P. Carvalho[2]

[1] UENF-Universidade Estadual do Norte Fluminense
Av. Alberto Lamengo, 2000, 28015-620, Campos dos Goytacazes, RJ, Brasil
{rivera, vaniave}@uenf.br

[2] IMPA-Instituto de Matemática Pura e Aplicada
Estrada Dona Castorina, 110, 22460 Rio de Janeiro, RJ, Brasil
pcezar@visgraf.impa.br

## ABSTRACT

Interference detection of arbitrary geometric objects is not a trivial task due to the heavy computational load imposed by implementation issues. The hierarchically structured bounding boxes help us to quickly isolate the contour of segments in interference. In this paper, a new approach is introduced to treat the interference detection problem involving the representation of arbitrary shaped objects. Our proposed method relies upon searching for the best possible way to represent contours by means of hierarchically structured rectangular oriented bounding boxes. This technique handles 2D objects boundaries defined by closed B-spline curves with roughness details. Each oriented box is adapted and fitted to the segments of the contour using second order statistical indicators from some elements of the segments of the object contour in a multiresolution framework. Our method is efficient and robust when it comes to 2D animations in real time. It can deal with smooth curves and polygonal approximations as well results are present to illustrate the performance of the new method.

## 1. INTRODUCTION

Many applications in Computer Graphics and Robotics demand real time analyses of interferences between objects. This type of analysis is used in animations, simulations, path planning to prevent interpenetration of objects in virtual environment [GoLiMa96], and self-interference in modelling [GrinsSchro01, VoliThalm94]. It is also recommended for use in optimization of two-dimensional stock cutting by means of compacting techniques [Milenko96] where the piecewise contours of the parts are objects of irregular geometry.

In animation, there are two strategies aiming interference detection: the structured approach and the direct approach. The structured approach requires additional storage space to allocate hierarchical structures, while optimizing the time consumed in the interference detection process. In the direct approach, the geometrical attributes of the objects, such as vertices, edges and faces, are main elements used to verify their neighborhood intersection. The direct approach does not use additional storage space, but in some cases it leads to lengthy geometric comparisons and its use in real time applications is not practical.

There are many efficient interference detection algorithms for three-dimensional objects that could be used in a two-dimensional context. However, they are not efficient in dealing with objects of arbitrary contours, specially if their contours present disturbance details[1] or have heterogeneities.

For instance, the incremental algorithm [Lin94] detects interferences using the Voronoi spaces defined by vertices, edges and faces of objects. The interference detection scheme based on the clipping methods [Hahn88, MoorWhil88, Kamat93] uses projections of the polygonal attributes of the objects. The witness technique [Baraff92] uses separation planes defined by faces, edges and vertices of convex objects. The tree sphere [Hubbard95] and the oriented box tree [GoLiMa96] methods define, respectively, a hierarchical bounding sphere and rectangular box, from the vertices and edges of polygonal contours of objects.

A hierarchical structure of envelopes represented in a binary, quaternary or octal tree manner, allows us to quickly discard the parts of the objects that are not in interference. In order to optimize the interference test time, the envelopes must enclose in adapted and adjusted fashion the segments of the object boundary and its details. Envelopes such as spheres, isothetic boxes[2] and ellipses do not necessarily optimize the time consumed in interference detection, because the minimum sphere enclosing a segment of a contour would contain more empty space than an ellipse covering the same segment. This happens because they are not adapted to the segments of the object contours. An oriented box might enclose a segment more tightly than an ellipse or a sphere (see Figure 1), because the orientation of the boxes is adapted to each segment of object contour. However, the intersection test for two spheres is faster than the intersection test for two oriented boxes or two ellipses. Nevertheless, this inconvenience is counter-balanced by the small number of boxes that have to be tested in order to identify interference. This yields a faster overlapping detection and, thus, fully justifies the use of oriented boxes over spheres (Figure 1 (d)).

Hierarchical structure approaches are also widely used in image synthesis techniques to minimize the testing time of complex scene geometries. Among several applications, we can mention ray-tracing [ArvoKirk89, RubiWhil8, GolsSalm87, Glassner84], radiosity [HanSalz91] and shadow [ChinFeiner89].

Bounding box trees were employed in ray-tracing and scene modelling [ArvoKirk89, RubiWhil80, WeHoGre84] and their use in computer animation was extended by Gottschalk et al. [GoLiMa96] for interference detection of three-dimensional objects with polygonal contours. In this work, we reframe these ideas, adapting them to interference detection in simulation of two-dimensional rigid object collision, where the details representing roughness are present in [RiCaVe01]. The orientations of the boxes are computed by sampling the respective bounded contour segments. We also introduce another way of computing the best fitting boxes according to a multiresolution framework which handles object contours from a coarse-to-fine two-dimensional perspective.

---

[1] Irregularities of the detail that define the roughness of the contour.

[2] Box of edges parallel bars to the coordinate system of the universe

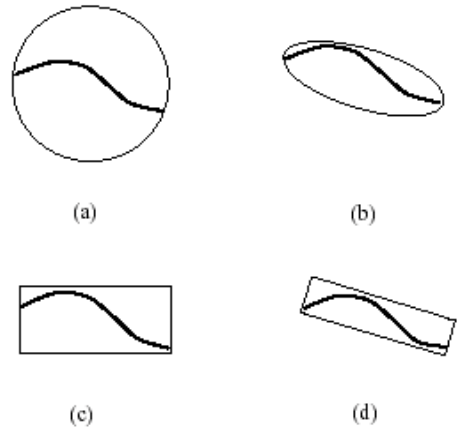

**Figure 1: Types of envelopes: sphere, ellipse, isothetic box and oriented box.**

The rest of this article is organized as follows. In Section 2, we define the object model used to state the proposed scheme. In Section 3, we study hierarchical structures, and, in particular, the generation of oriented bounded boxes. In Section 4, we introduce our model for interference detection of objects in animation and in Section 5 we present the implementation results of our method. Concluding remarks are given in Section 6.

## 2. OBJECT OF ARBITRARY GEOMETRY CONTOUR

For the purpose of validation of our model, we consider two-dimensional objects with contours defined by periodic and continuous cubical B-spline curves [RogAda90]. We define an object by $m$ control points $C = \{c_i\}_{i=0,...,m}$ that generate $m$ segments of the curve, $\{f_i\}_{i=0,...,m}$. The contour of an object is given by $\cong \bigcup_{i=0}^{m} f_i$. We allow the existence of roughness (or noise) and this possibility is accounted for by means of a normal probability density function

$N(0,\omega_i)$, which contains noise in each segment. The variance $\omega_i$ represents the amount of details present in $f_i$. Each segment is defined geometrically and the roughness details might be reproduced probabilistically using a tolerance interval $\varsigma_j = q\omega_j$ with $0 \le q \le 1$ [Morrison76]. Figure 2 shows the segment $f_j$ with roughness details inside of an oriented box. The details of $f_j$ are limited by the tolerance interval $\varsigma_j$.

For the purpose of multiresolution representation, $f$ is denoted by $f^n$ where $n = \log_2 m$, and its respective points of control by $C^n$ [GomVel98, RiCaVe99]. Therefore, if $f^n$ defines a contour at a given geometrically defined, then $f^{n-1}$ is a coarser representation of $f^n$, where the detail of this contour is $g^{n-1} = f^{n-1} - f^n$. In the multiresolution representation, $f^n$ can be expressed as a sequence of coarse resolutions $f^{n-1}$, $f^{n-2}$, ..., $f^{min}$, where $f^j = f^{j-1} + g^{j-1}$, $min \le j \le n$, and *mim* is the minimum number of control points that permit to define a closed object in a multiscale representation. In this work, we use the bi-orthogonal wavelet transform in its fast version [GomVel98] in order to handle contours more efficiently.

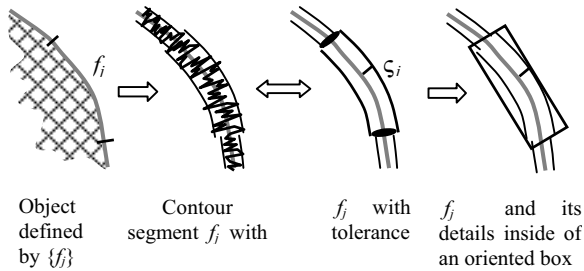

| Object defined by {$f_j$} | Contour segment $f_j$ with | $f_j$ with tolerance | $f_j$ and its details inside of an oriented box |

**Figure 2: A segment of a contour with details inside an oriented box.**

In practice, the transformations are performed at the level of the control points of the contour. Thus, $C^n$ is transformed into $C^{n-1}$, this into $C^{n-2}$ and so forth, until level $C^{min}$, using analysis filters $A^j$ and $B^j$ such $C^{j-1} = A^j C^j$ and $D^{j-1} = B^j C^j$, for $min \le j \le n$. The vector $D^{j-1}$ is the detail coefficient vector, which is not used in this work. Instead, the scale coefficient vector $C^{j-1}$ is used to get the coarse version $f^{j-1}$ such $f^{j-1} = C_{j-1} A^{j-1}$, where $A^{j-1}$ is the cubic B-spline basis function vector.

## 3. HIERARCHICALLY ORIENTED BOUNDING BOXES

If each segment $f_i^n$ and its details are bounded by an oriented box $b_i^n$, all contours of the object will be covered by a set of boxes called *elementary boxes*. In order to construct a binary tree of oriented boxes, each pair of adjacent elementary boxes, denoted by $b_{2i}^n$ and $b_{2i+1}^n$, is tightly enclosed by another box $b_i^{n-1}$ called *super box*. The elementary boxes contained by each adjacent pair of super boxes $b_{2i}^{n-1}$ and $b_{2i+1}^{n-1}$ are covered in an adapted and adjusted form by another super box $b_i^{n-2}$. Following this process, we construct a binary tree, where the root is a super box $b_1^0$ which bounds in an adapted and adjusted fashion all elementary boxes of the object. The inclusion of elementary boxes inside of the superior boxes allows the conservation of the characteristics of the segment such as their roughness details among other features. Figure 3 shows one segment of the oriented bounding box tree.

The orientation of a box is better characterized by its main axis orientation, and it is defined by the behavior of the contour segment bounded by this box. The best form to determine the behavior of the segments is to use the covariance matrix O computed with the elements related to these segments, as detailed in the next subsection. For the two-dimensional case, the unitary eigenvectors $\mathbf{e}_1$ and $\mathbf{e}_2$ of O are used to represent the main axes.

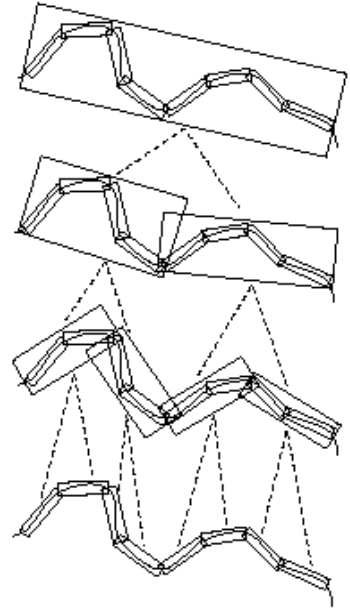

**Figure 3: A segment of a hierarchically structured oriented bounding boxes.**

In general, the definition of an oriented bounding box is done in two steps: *adaptation* and *adjustment*. The adaptation step consists of computing the orientation of the box. The adjustment step calculates the box dimensions. While defining elementary boxes, if we have contours presenting more complex details, such as too much roughness for instance, then a third step called *increment* is performed to include these details in the box and better characterize the segments.

The information that defines a box $b_i^n$, that is the unitary main axes, the dimensions, center point, tolerance $\varsigma_i$ and the length of the segment $f_i^n$ must be stored in the data structure that better describes the oriented box. This information will be useful in the calculations of the respective super boxes, and the interference test.

## Elementary box

The elementary box $b_i^n$ is generated from the elementary segment $f_i^n$ of the object contour and its disturbances (roughness) in the adaptation, adjustment and increment steps.

**Adaptation**: The covariance matrix $O \subset R^{2 \times 2}$ of the centroid is computed using the simple average $\sigma$ of the $r$ points $\mathbf{p}_i = (p_i^x, p_i^y)$ uniformly sampled on $f_i^n$, where $p_i^x$ and $p_i^y$ are the coordinates $x$ and $y$ of $\mathbf{p}_i$. We considered $r = 5$ as appropriate to compute the orientation of the adapted axis according to the alignment of the segment, because a bigger concentration of points in some part of $f_i^n$ cannot reflect in the axis the real tendency of the segment. Figure 4 shows an elementary box defined by uniform sampling of one type of curved segment. The elements of the covariance matrix, in this case for $x=1, 2$ and $y=1,2$, are of the form:

$$\omega_{xy} = \frac{1}{r-1}\sum_{i=1}^{r}(p_i^x - \sigma^x)(p_i^y - \sigma^y)$$

with $\sigma = (\sigma^x, \sigma^y)^T = \frac{1}{r}\sum_{i=1}^{r}(p_i^x, p_i^y)^T$.

**Adjustment**: The segment $f_i^n$ is projected on the axis with origin in $\sigma$. The sides of the box are defined by the segments of bigger dimension between the projections on each axis.

**Increment**: This step is used when the contour of the object has some roughness characterized by $N(0, \omega_i)$, where $\omega_i = \varsigma_i / q$ for $0 < q < 1$. The sides of the box are added with the tolerance $\varsigma_i$ that represents the quota, near to the superior, of the details associated to the segment $f_i^n$. After define the dimensions of the box $b_i^n$, we recompute their new center point.

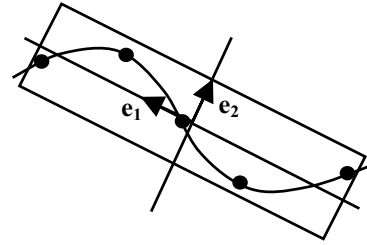

**Figure 4: Elementary boxes adjustment with *m*=5 sampling points.**

## Super boxes

The super boxes contain two or more elementary boxes; therefore, the adaptation and the adjustment steps are computed having in mind other considerations besides the ones used to compute the elementary boxes. We wish to construct a super box $B$ tightly fitting a contour and incorporating $k$ adjacent elementary boxes.

### 3.1.1 Adaptation.

The best adapted box to a segment of object contour depends solely on the choice of its main axis, better saying on the covariance matrix $O$ that defines the unitary axes. In this work, we propose a method to compute the segment orientation in a multiresolution fashion. First, in order to do that, it is necessary to understand how to determine orientation when dealing with elementary boxes [Rivera00]. Our proposed method generates more suitable segment orientations of the contours than when only elementary boxes are used, but it requires a previous contour decomposition process into lower resolutions. The computational load of this extra step is negligible.

### 3.1.1.1 Adaptation based in elementary boxes

The covariance matrix is computed by using the centroids of the elementary boxes weighed by the length of the corresponding segment. So, it is not influenced by the concentration of small segments in any sector of the contour.

If $\mathbf{p}_i = (p_{i,x}, p_{i,y})$ is the centroid of the box $b_i^n$, and $l_i$ is the arc length of the elementary segment $f_i$, the average of $k$ segments is given by

$$\sigma = \frac{1}{l}\sum_{i=1}^{k} l_i \mathbf{p}_i \quad \text{with} \quad l = \sum_{i=1}^{k} l_i$$

An element $\omega_{xy}$ of the covariance matrix $O$, is calculated, in this case, as follows:

$$\omega_{xy} = \frac{1}{l}\sum_{i=1}^{k} l_i (p_i^x - \sigma^x)(p_i^y - \sigma^y)$$

*3.1.1.2 Adaptation based in mult-iresolution*

Using the wavelet transform in a multiresolution framework, the correspondence between the segments of two adjacent resolutions of the object contour is established by following theorem stated in [Rivera00].

**Theorem 3.1**: *Given the adjacent inferior resolutions $f^j$ and $f^{j+1}$ of the contour $f^n$, for $j < n$, then the segment $f_i^{j+1}$ converges in $f_{2i}^j$; $f_{2i+1}^j$.*

The previous theorem allows us to merge pairs of segments and their respective boxes which results in a box belonging to the immediate superior resolution (super box). Moreover, the version $f^j$, for $\min \leq j < n$, represents the average of the version $f^{j+1}$, according to the theory of multiresolution based on the wavelet transform [GomVel98]. Each segment of $f^j$ defines the orientations of the boxes of level $j$ in the hierarchical structure. For example, the orientation of the box $b_i^{n+1}$ is defined by the segment $f_i^{n+1}$, and its dimensions are defined by the elementary boxes $b_{2i}^n$ and $b_{2i+1}^n$ that bound segments $f_{2i}^n$ and $f_{2i+1}^n$, respectively. These elementary segments oscillate around $f_i^{n+1}$ [Rivera00]. Figure 5 illustrates one super box defined by a segment $f_r^{n+1}$ and their two corresponding elementary boxes.

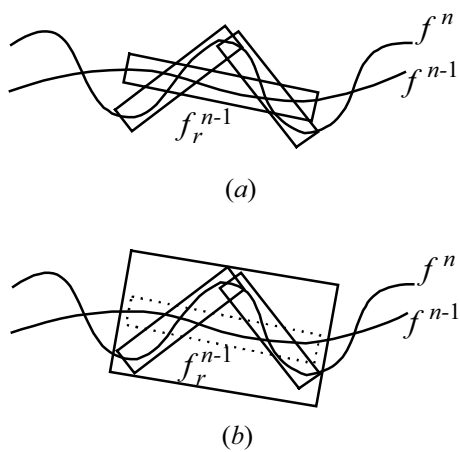

**Figure 5: Super box based in multiresolution segments: (a) adaptation; (b) adjustment.**

The main axis of the super box $b_i^j$ is defined from the covariance matrix $\overline{O}_i^j$ computed similarly to the adaptation step of the elementary box method, by uniform sampling the segment $f_i^j$. So, the orientations of the boxes of level $j$ are defined by the contour segment of $f^j$ that is lower version of $f^n$ in multiresolution representation, for $\min \leq j \leq n$. The other levels of boxes, $b_i^j$ for $0 \leq j < \min$, are computed by using the adaptation based in elementary boxes, because we can only define lower resolutions of $f^n$ until $f^{\min}$.

*3.1.2 Adjustment*

The vertices of the $k$ elementary boxes are projected over the axes $\mathbf{e}_1$ and $\mathbf{e}_2$ having $\sigma$ as the origin of the coordinate system. The box dimensions are defined by the largest projection segments on each axis among the $k$ elementary boxes. Once these projections are known, we compute the vertices and the accurate centroid of $B$.

The method of adaptation in multiresolution representation permits us to compute the main axis of super boxes better adapted to the segment of corresponding resolution, in such way the boxes are adapted to the respective segments of $f^n$. In Figure 6, we can observe the differences between the super boxes defined by the elementary boxes method (a), and the ones defined by the multiresolution representation (b). The first method does not generate good boxes when compared to the second one. Moreover, we can observe that the areas of the boxes generated by the multiresolution representation are smaller than similar boxes gotten by the method of adaptation based on elementary boxes.

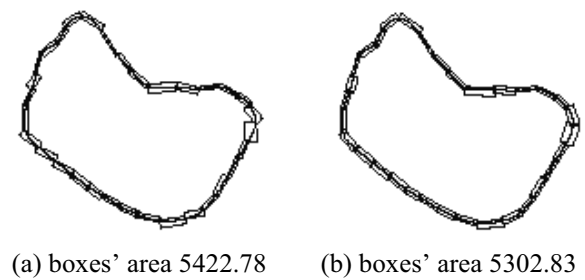

(a) boxes' area 5422.78     (b) boxes' area 5302.83

**Figure 6: Boxes for level 4 of the tree: (a) representation obtained with elementary boxes; (b) representation obtained using segments in multiresolution.**

## 4. INTERFERENCE DETECTION

If two objects are in interference, then some of their elementary boxes are overlapping. The interference analysis, formulated by Gottschalk et al. [GoLiMa96], which is made recursively starting from the roots of the trees through the children while the intersection between boxes is detected. When finishing, in case that it has registered elementary boxes, it means that it is possible the interference of two objects. The intersection of the segments enclosed by the boxes in interference indicates the interference of the respective objects. The intersection of the segments will be verified numerically, considering the details to define the possible situation between the two objects. Figure 7 shows two possible situations when two objects are present in a scene: separated or in interference.

Two boxes do not intersect if and only if there is a line such that the orthogonal projections of the boxes onto that line do not intersect. Actually, it suffices to test the lines which are parallel to a side of one of the boxes, which takes constant time.

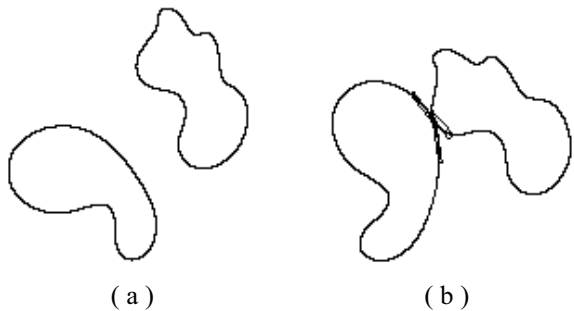

**Figure 7: Situations: (a) separate; (b) in interference.**

## 5. EXPERIMENTAL RESULTS

Our method was implemented using the "C" language. The graphical libraries IUP/LED and CD were employed for visualization purposes. These package are a courtesy of the Group of Graphical Technology (TecGraf) from Pontifícia Universidade Católica do Rio de Janeiro (PUC-Rio).

### Performance of the interference detection

The efficiency of interference detection in animations using the method of hierarchical bounding boxes for complex objects with disturbances is inspired by the OBBtree approach [GoLiMa96]. To illustrate the performance of our method, examples involving three situations with three objects, where each one has 512 segments, are formulated. The situations are shown in Figure 8. Each experiment consists of dealing with 3069 oriented boxes distributed over three trees. The implementation was done in a Sun SPARCstation 20 with Solaris. The measured time is the minimum time in seconds that takes to run an experiment in this multitask environment.

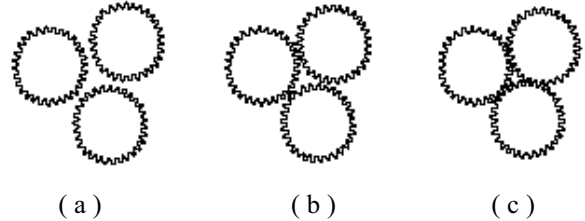

**Figure 8: Objects with 512 segments each: (a) non-interfering objects; (b) contact between two objects; (c) there are multiple contact points among all objects.**

Table 1 shows some results. It is observed that, exactly when there is interpenetration of objects, only one fraction of the boxes is compared (459 of the 3069) and the processing time is of the order of 0.01 seconds.

| Number of objects: | | 3 | |
|---|---|---|---|
| Total number of segments: | | 1536 | |
| Total number of boxes of the trees: | | 3069 | |
| Situation | Time in seconds | Number of boxes tested | |
| | | Elem. boxes | Multiresolution |
| (a) | 0.0001 | 35 | 35 |
| (b) | 0.003 | 172 | 170 |
| (c) | 0.01 | 459 | 457 |

**Table 1: Numerical results of the situation corresponding to Figure 8.**

### Examples of results in animations

Results for some frames of sequences of animations of plain complex objects. In Table 1, we can observe that the interference detection with the new method is similar or better than the one obtained with the elementary box based method. Figure 9 shows two frames of multiple object animations, and Table 2 shows the numerical results when testing for interferences. We can observe that in any situation the boxes obtained by our multiresolution scheme are better than the ones obtained with the other method. Similar situation can be observed in Figure 10 which is related to Table 3.

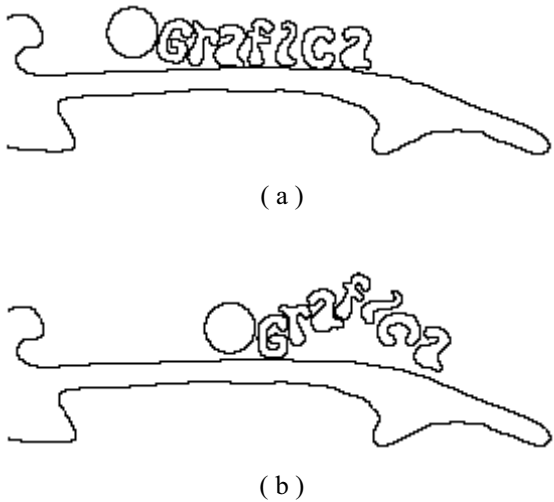

( a )

( b )

**Figure 9: Two frames of an animated sequence.**

| Number of objects: | 9 | |
|---|---|---|
| Total number of segments: | 615 | |
| Total number of boxes of the trees: | 1225 | |
| Situation | Number of boxes tested | |
| | Elem. boxes | Multiresolution |
| (a) | 523 | 520 |
| (b) | 734 | 732 |

**Table 2: Numerical results for the situation shown in Figure 9.**

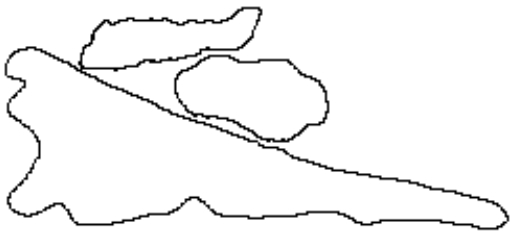

**Figure 10: Irregular geometric objects in contact with tolerance.**

| | |
|---|---|
| Number of trees: | 3 |
| Total number of segments: | 1536 |
| Total number of boxes build: | 3075 |
| Boxes based in elementary boxes tested: | 229 |
| Boxes based in multiresolution tested: | 223 |

**Table 3: Numerical results for the situation shown in Figure 10.**

## 6. CONCLUSIONS AND FUTURE WORKS

Hierarchically structured bounding boxes permit us to quickly test for interference of complex objects in movement. So, we can isolate the pieces of a contour that may possibly be in contact with another contour. Later, the verification and the calculations of contact points are made by local procedures on the elementary segments of the object contours. In real time application, where the test for interferences is made in each time step, the quick isolation of segments of object contours is very important for fast contact point detection. The isolation process must test a few pairs of bounding boxes, and must use few arithmetic operations in order to minimize the time consumed in the contact detection process. So, we consider that our method is the best alternative to interference detection in animation and simulation when it comes to manipulation of objects of complex geometries.

Better saying, our method is efficient and robust for two-dimensional animations in real time. Furthermore, it is one of a few approaches that can handle smooth curves and polygonal approximations as well. Most researchers restrict their models to this last case.

An application of the model formulated here we want to apply in the cutting stock problem, where each object could represent a desired piece of an entire material that has to be cut off, such as leather for shoes, blades, cards, etc. The objects compact following physical laws. Adding some heuristics to allow to set objects into motion until reaching equilibrium. Another application of fast interference detection is the modelling of curves and surfaces for direct manipulation with self-interpenetration restrictions between different segments of the same object.